\documentstyle[psfig,conf_iap]{article}

\def\lya{Ly$\alpha$~}
\def\kms{km~s$^{-1}$}

\begin{document}

\heading{Heavy element enrichment in the IGM at high redshift}

\par\medskip\noindent

\author{S.~Savaglio}
\address{European Southern Observatory, Schwarzschildstr. 2,
        D--85748 Garching, Germany}

\begin{abstract}
We present a detailed analysis of the ionisation
state and heavy element abundances in the Intergalactic Medium
(IGM). The CIV doublet is shown by 30 \% of the 
182 selected optically thin \lya clouds
in 10 QSO lines of sight. Direct
metallicy calculations have been performed on individual systems with
detected CIV and SiIV (10\% of the sample) varying the
UV photoionising source, cloud density and size and silicon relative
abundance. The best solutions for carbon content in this subsample 
(redshift coverage $z=2.6 - 3.8$) span between 1/6 and 1/300
of the solar value with no evidence of redshift evolution in both the
metallicity and the ionising source. Global properties of the
whole sample indicate that the metallicity in \lya clouds with CIV and SiIV
is not typical of the IGM. The redshift evolution of the UVB
is one of the possible
sources of the observed SiIV/CIV trend presented by Cowie and collaborators
during this meeting. Future detection of heavy elements in lower HI
column density ($\log N_{\rm HI} < 14.5$)
\lya clouds relies on the presence of OVI and NV at $z=1-2.5$.

\end{abstract}

\section{Introduction}

For many years the  \lya forest has been
considered a different class of objects with respect to galaxies.
The available 
sensitivity was too low to detect any sign of non--primordial
composition in the intergalactic gas clouds at  high redshift.
Thanks to the advent of high resolution and signal--to--noise
spectroscopy, the old idea on the majority of quasar absorption lines 
has been 
revisited and opened in the last few years a still pending debate
on the connection between the \lya forest and the
galaxy formation of the early Universe. The detection of ions
different from CIV in optically thin
\lya clouds 
is made complicated by harder
observational conditions, whereas the still too poor knowledge of the
ionisation mechanisms which determine the ion abundances in those
clouds has often discouraged attempts of
metal content estimations as a function of
redshift and of HI column density.
However abundance investigation of the \lya clouds has fundamental
implications in the understanding of the enrichment processes in the IGM
by Pop III stars in the $z>5$ Universe.

\section{Metallicity measurements}

\begin{figure}
\centerline{\vbox{
\psfig{figure=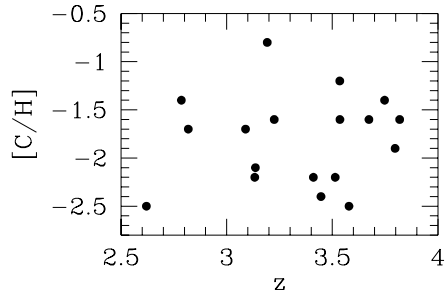,height=4cm,angle=0}
}}
\caption[]{\label{f0} Carbon abundance relative to solar as a function
of redshift in the
optically thin \lya lines which both show CIV and
SiIV doublets.}
\end{figure}

The sample of optically thin
absorption lines with  $14.5 < \log N_{\rm HI} < 16.5$ 
has been obtained by high resolution spectroscopy, mainly HIRAS/Keck
(Songaila 1997b) but also by EMMI/NTT for the $z\gsim 3.7$ systems 
(Savaglio et al. 1997). For all the systems  CIV and/or SiIV and CII
detections or upper limits are given in redshift coverage
$2 < z < 4.5$. 
The lower bound in $N_{\rm HI}$ 
is due to the very rare metal detection in lower
column density systems. In this range even if the line can
be saturated (depending on the Doppler width)
 Monte Carlo simulations showed that fitting procedures
of synthetic individual lines with similar resolution and S/N ratio
of the observed spectra give HI column density errors which are less
than a few tens of $dex$
(for $b=25$ \kms, $\log N_{\rm HI} = 15.5$, FWHM = 12 \kms
and S/N = 20 this is typically 0.1 $dex$). The blending effect has a much
more dramatic impact on column density uncertainties
and for this reason, we consider in the
case of complex structures as an individual cloud 
the total column densities of HI and of
metal lines.

Estimating the heavy element  content in the \lya clouds is mostly
complicated by the poor knowledge of the ionising sources. As a first
simplification, we assume that this is dominated by photoionisation
of the UV background and neglect any other
mechanism. 
Collisional ionisation is important when the gas temperature 
exceeds $10^5$ K. At that temperature, the
Doppler parameter for HI is 41 \kms, well above the mean value
typically found
in \lya clouds. The analysis of metal lines in \lya clouds (Rauch
et al., 1997) shows that the mean ``Doppler'' temperature in these clouds is
$\sim4\times10^4$ K, making any evidence of collisional ionisation
hard to justify. 
Once the photoionisation equilibrium is assumed, we first
consider the subsample of \lya clouds which show both CIV and SiIV
absorption. To calculate the metallicity we use CLOUDY and assume six
different shapes for the UV background normalized to the value at
the Lyman limit ($J_{912} = 5\times10^{-22}$ erg s$^{-1}$ cm$^{-2}$
Hz$^{-1}$ sr$^{-1}$) changing the parameter $S_L =
J_{912}/J_{228}$ in the range $200-3000$. We varied 
the [C/H] and gas density in such a
way to reproduce the observed CIV. We also assume the
relative silicon--to--carbon abundance to be between 0 and three times solar
and consider the cloud size along the line of sight to be in the
range 1 kpc $\lsim R \lsim 50$ kpc. Given these assumptions, we
obtain for this subsample a set of 18 [C/H] measurements shown in
Fig.~\ref{f0}.  Carbon abundance in clouds with detected
carbon and silicon  has a large spread with
mean values of [C/H] $= -1.8$ and no evidence of redshift evolution.  We
notice that this sample might consist of metal--rich \lya clouds
since it has been selected because of the SiIV detection 
and might not be representative of the whole population of \lya clouds. 
In a recent work, Songaila (1997a) 
has estimated the total universal metallicity
at $z\sim3$ (assuming that at that time the baryonic matter of the
Universe mostly resides in the \lya forest)
to be in the range 1/2000 and 1/630 relative to solar.

\begin{figure}
\centerline{\vbox{
\psfig{figure=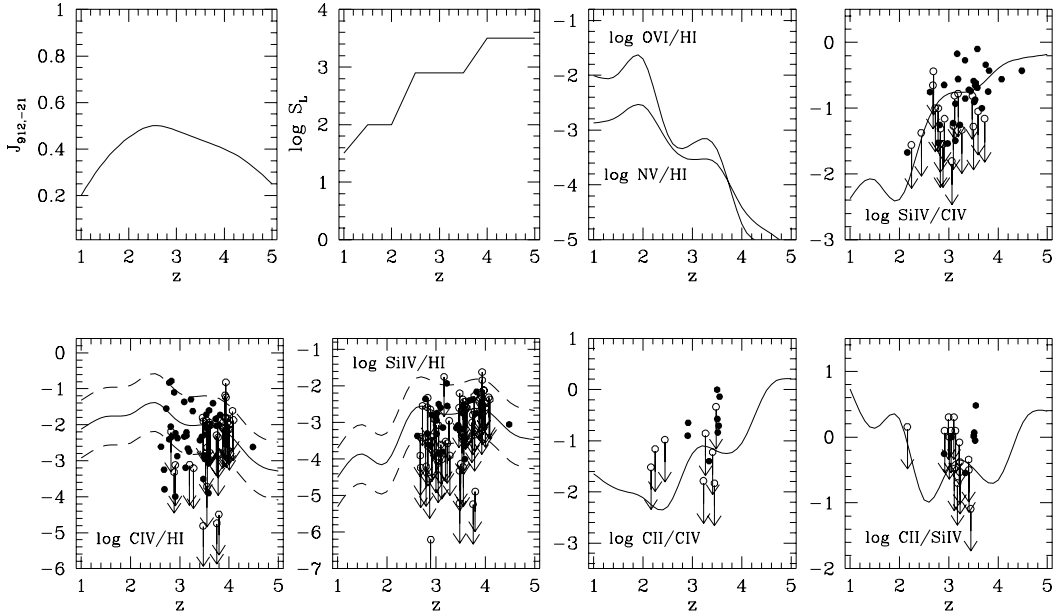,height=8.5cm,angle=0}
}}
\caption[]{\label{f1} Ion column density ratios as a function of
redshift. Solid curves are models assuming the UVB 
as described by the first two panels ($\log N_{\rm HI} =
15$ and [C/H] $=-1.8$). Dashed curves are the same but for [C/H] 
$=-1.8\pm0.8$.}
\end{figure}

\begin{figure}
\centerline{\vbox{
\psfig{figure=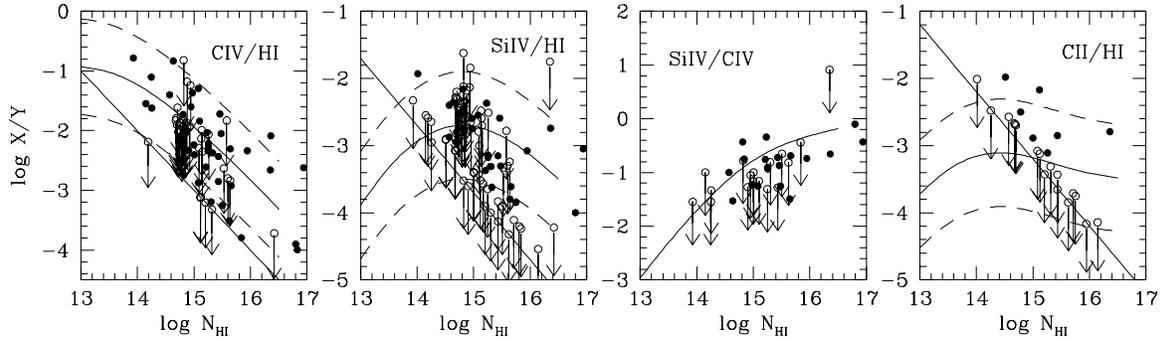,height=4.9cm,angle=0}
}}
\caption[]{\label{f2} Ion column density ratios as a function of
HI column density. Solid and dashed curves 
are model calculations assuming the UVB
at $z=3$ of Fig.~\ref{f1}  ([C/H] $=-1.8\pm0.8$). Straight lines
represent detection limits.}
\end{figure}

In a different approach, we consider the whole sample
and regard the global observed properties instead of the individual
 systems and compare with models. Results of column density ratios on the $z$ and
$N_{\rm HI}$ planes are shown in Figs.~\ref{f1} and
\ref{f2}. In Fig.~2 we investigate the redshift evolution of
observed column densities in the case of $S_L$ and $J_{912}$ as reported.
The discussed trend of SiIV/CIV (Cowie et al., this
conference proceedings) can be
reproduced by a redshift evolution of $S_L$ from 200 at $z\sim2$ to 3000
at $z\sim4$. The same model can take into account other observed ion ratios.
In Fig.~3 we compare observations with CLOUDY
models assuming that all the clouds of the sample are at the same mean
redshift of $z=3$ with $S_L=800$ 
and the gas density proportional to the square root of
$N_{\rm HI}$, as given in the case of spherical clouds in
photoionisation equilibrium with the UVB. In both figures the solid
lines are obtained for metallicity [C/H] $= -1.8$ and [Si/C] = [O/C] =
0.5, [N/C] = 0. Models of photoionisation equilibrium 
can include the majority of metal
detections (also considering the metallicity spread) but
CII/HI which, as function of $N_{\rm HI}$, looks to be
steeper than calculated. Additional observations of CII would probably
cast further light on the discussion on the ionisation state and metal content
in the \lya clouds. 
In both figures, the numerous upper limits falling below the dashed curve
[C/H] $= -2.6$ is an indication that in many clouds the
metallicity is lower than the values found in the selected sample.

\section{The future}

The investigation of low and
intermediate redshift ($z=1-2.5$) 
observations of OVI and NV in $\log N_{\rm HI}\lsim 14$
\lya clouds might succeed in answering the question of how efficient the mixing
processes
in the IGM at high redshift has been. Relative abundances can
provide new hints on the study of metal production by Pop III
stars. In particular NV since it has been predicted to be underproduced in
massive stars with low initial metallicity (Arnett 1995).
More observations of the SiIV/CIV ratio for $z<2$
and $z>4$ are a challenging probe of the redshift evolution of the UVB,
though this can be one of the many possible reasons for the
observed SiIV/CIV trend (another would be redshift evolution
of the gas density being lower at lower redshift). 
More interesting conclusions await outcomes
from new high quality data of Keck observations.

\end{document}